\documentclass[
  fontsize=12pt,
  paper=a4,
  DIV=14, 
]{scrartcl}

\usepackage{graphicx,xcolor}
\usepackage{amsfonts,amssymb,amsbsy,amsmath,amsthm}
\usepackage{txfonts}
\usepackage{paralist}
\usepackage{physics}
\usepackage{siunitx}
\usepackage[colorlinks,linkcolor=black,citecolor=black,urlcolor=black,breaklinks]{hyperref}

\addtokomafont{captionlabel}{\bfseries}
\setcapwidth[c]{0.95\textwidth}

\AtBeginDocument{}

\KOMAoptions{
  headings=standardclasses,%
}

\usepackage{nicefrac}

\numberwithin{equation}{section}

\makeatletter
\renewcommand{\@biblabel}[1]{#1\hfill \hspace{-0.2cm}}
\makeatother

\title{Dynamic fracture with continuum-kinematics-based peridynamics}

\author{%
  Kai Friebertshäuser$^{a*}$, 
  Christian Wieners$^b$, 
  and
  Kerstin Weinberg$^a$
}

\date{\normalsize%
\begin{compactitem}
  \item[$^a$] Chair of Solid Mechanics, University of Siegen, Siegen, Germany%
  \item[$^b$] Institute of Applied and Numerical Mathematics, KIT, Karlsruhe, Germany
\end{compactitem}%
\vspace{3mm}
\begin{compactitem}
  \item[$^*$] Correspondence: kai.friebertshaeuser@uni-siegen.de; Tel: +49 0271 740 2185%
\end{compactitem}%
}

\begin{document}

\maketitle
%


%
\vspace*{-9mm}
\begin{abstract}
This contribution presents a concept to dynamic fracture with continuum-ki\-ne\-ma\-tics-based peridynamics. Continuum-kinematics-based peridynamics is a geometrically exact  formulation of peridynamics, which adds surface- or volumetric-based interactions to the classical peridynamic bonds, thus  capturing the finite deformation kinematics correctly. The surfaces and volumes considered for these non-local interactions are constructed using the point families derived from the material points' horizon.

For fracture, the classical bond-stretch damage approach  is not sufficient in con\-ti\-nu\-um-kinematics-based peridynamics. Here it is 
extended to the surface- and volume-based interactions by additional failure variables considering the loss of strength
in the material points' internal force densities. 
By numerical examples, it is shown that the approach can correctly handle crack growth, impact damage, and spontaneous crack initiation under dynamic loading conditions with large deformations.
\end{abstract}

\bigskip
\noindent\textbf{Keywords:} dynamic fracture, peridynamics, continuum-kinematics-based peridynamics, crack propagation, impact


\section{Introduction}

Predicting  crack propagation and material damage is still very challenging in computational mechanics.
Fracture problems have been addressed by various computational methods such as damage models or 
discontinuous finite element discretizations 
\cite{XuNeedleman1994,OrtizPandolfi1999,DallyBilgenWernerWeinberg2020} 
and  phase-field fracture simulations \cite{MieheMaute2016,WilsonLandis2016,BilgenWeinberg2021}.
All these approaches are based on the classical continuum mechanics assumption of a homogeneous bulk material.
Peridynamics allows an alternative approach to fracture because it models the material in a non-local form. 
Initially introduced by Silling \cite{silling2000,silling2005}, peridynamics uses integral equations to describe the relative displacements and forces between material points.
Concepts like stress and strain are absent, and the behavior of a material point is described solely by its interactions with other material points.

The original peridynamic concepts were restricted to the interaction of bonds, which has limited the ability to account for a material's Poisson ratio other than $\nicefrac{1}{4}$ for 3D problems. Other formulations,  like ordinary state-based peridynamics  and non-ordinary state-based peridynamics, address this problem, cf.~
A  new approach was recently introduced by Javili, McBride \& Steinmann, who propose a continuum-kinematics-based reformulation of peridynamics \cite{Javili2019, Javili2020, Javili2021}. This geometrically exact formulation relies on an analogy to the classical continuum mechanics and is intrinsically designed to capture the lateral contraction of the material correctly.
Three  types of material point interactions are introduced, namely bond-, surface- and volumetric-based interactions, which correspond to the invariants of a general deformation. 
The relationships between the material parameters of continuum-kinematics-based peridynamics and isotropic linear elasticity were recently  elaborated  for two- and three-dimensional problems, \cite{Ekiz2021, Ekiz2022}.

%

The new kinematics of continuum-kinematics-based peridynamics requires a new concept of damage and fracture. Because of the three different types of interactions, it is no longer sufficient to understand material damage as a bond-based phenomenon 
Thus we enrich the material description by  kinematic variables that account for the loss of load-carrying capacity in the material's  force density, which is also extended by a
density related to contact.  In that way, crack nucleation  and propagation as well as impact damage can be modeled. 
To the knowledge of the authors, this is the first concept of damage within the  continuum-kinematics-based peridynamics framework.

This manuscript is organized as follows. In Section 2 the notation and the necessary theory of  continuum-kinematics-based peridynamics are introduced. Here the one-, two-,, and three-neighbor interactions are defined, and  the model is extended to contact of two or more bodies. 
Section 3 is the  paper's core; here we introduce our damage model. Sections 4-6 present numerical examples. We start with a model validation by a simple 2d crack growth for a mode I tension test in  Section 4. Next, the crack initiation due to reflected impact  waves is presented in Section 5 using the example of a curved bar \cite{WeinbergWieners2022}. Finally, in Section 6, continuum-kinematics-based peridynamics is used to compute damage induced by the impact of a sphere on a brittle plate.

\section{Theory of continuum-kinematics-based peridynamics}

In peridynamics, a body is considered as a set of $N$ points in Euclidean space $\mathbb{R}^3$, and the dynamics is described by the movement of these points.
%
%
Specifically, the bijective mapping
\begin{align}
  \vb*{\Phi}_t \colon &\mathbb{R}^3 \mapsto \mathbb{R}^3
  \\
  \nonumber
  & \mathcal{B}_0 \mapsto \mathcal{B}_t
  \qquad \exists \; \vb*{\Phi}_t^{-1} \colon \; \mathcal{B}_t \mapsto \mathcal{B}_0
\end{align}
describes the transformation (deformation) of the body from reference
configuration $\mathcal{B}_0$ to current configuration $\mathcal{B}_t$
at the time $t$.  The movement of a body can therefore be described
as a parametrical (temporal) sequence of deformations $\vb*{\Phi}\colon
\mathcal{B}_0 \times [0, T] \mapsto \mathbb{R}^3$.

\begin{figure}[ht]
\centering
\includegraphics[]{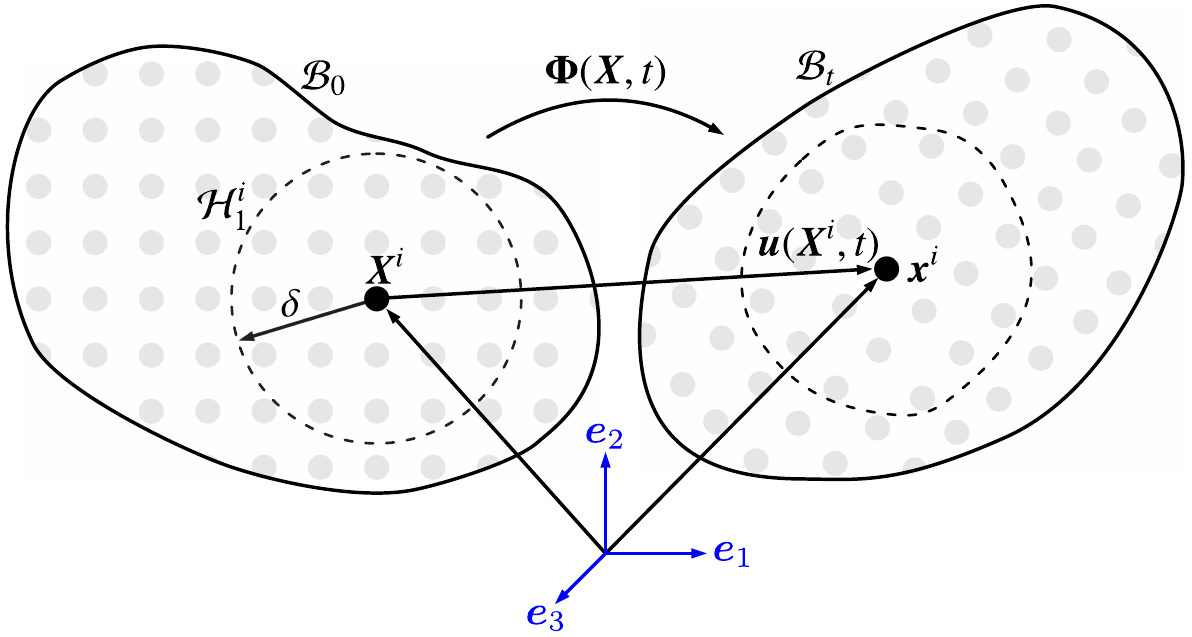}
\caption{Deformation of body $\mathcal{B}_0$ to $\mathcal{B}_t$ and $\vb*{X}^i$ to $\vb*{x}^i$ of the material point $i$}
\label{fig:pd_deformation}
\end{figure}

The point position in material configuration is described by $\vb*{X}^i$ and in current configuration as 
\begin{equation}
\vb*{x}^i = \vb*{X}^i + \vb*{u}(\vb*{X}^i, t) = \vb*{\Phi}_t(\vb*{X}^i)\; ,
\end{equation}
with the displacement vector $\vb*{u}(\vb*{X}^i, t)$ and $i=1,\dots,N$.
Points interact only with other points inside of their specified \emph{neighborhood} $\mathcal{H}_1^i$, which is defined as the set of points inside the spherical space with the radius $\delta \in \mathbb{R}^+$, also called the \emph{horizon} $\delta$ (see \autoref{fig:pd_deformation}).
Accordingly,  $\mathcal{H}_1^i$ includes all points $\vb*{X}^j$ inside the horizon of point $\vb*{X}^i$ in the reference configuration of the body $\mathcal{B}_0$,
\begin{equation}
  \mathcal{H}_1^i = \left\{ \vb*{X}^j  \in \mathcal{B}_0\; | \;  0 < \abs{\vb*{X}^j - \vb*{X}^i} \leq \delta \right\} \quad \forall \; \vb*{X}^i\in \mathcal{B}_0 \; .
\end{equation}

The equation of motion for point $i$ reads
\begin{equation}\label{eq:equation_of_motion}
\rho \, \vb*{\ddot{u}}(\vb*{X}^i, t) = \vb*{b}_0^{\mathrm{int}}(\vb*{X}^i, t) + \vb*{b}_0^{\mathrm{ext}}(\vb*{X}^i, t) \qquad \forall \; \vb*{X}^i \in \mathcal{B}_0, \; t \geq 0
\end{equation}
with the density $\rho$, the point acceleration vector $\vb*{\ddot{u}}$, and the point force density vectors $\vb*{b}_0^{\mathrm{int}}$ and $\vb*{b}_0^{\mathrm{ext}}$, which denote force per unit undeformed volume.
The external force density $\vb*{b}_0^{\mathrm{ext}}$ results from the external forces that are acting on the body and the internal force density $\vb*{b}_0^{\mathrm{int}}$ from the interactions between the individual material points.
Peridynamics can be understood as a Lagrangian particle method, because all equations are mapped to the reference configuration.
In the following, the notation $\vb*{u}^{i} = \vb*{u}(\vb*{X}^i, t)$ and $\vb*{b}_0^{\mathrm{int}, \, i} = \vb*{b}_0^{\mathrm{int}}(\vb*{X}^i, t)$ will be used for improved readability.

All our simulations consider a short period of time, therefore an explicit time integration scheme is used.
We employ the Velocity-Verlet algorithm of Littlewood \cite{littlewood2015}, where for each time $t$ and material point $i$, the acceleration, velocity and displacement are calculated as
\begin{subequations}
\begin{align}
\vb*{\dot{u}}^{i}(t+\tfrac{1}{2} {\vartriangle} t) &= \vb*{u}^{i}(t) + \frac{{\vartriangle} t}{2} \, \vb*{\ddot{u}}^{i}(t) \; ,\\
\vb*{u}^{i}(t+{\vartriangle} t) &= \vb*{u}^{i}(t) + {\vartriangle} t \, \vb*{\dot{u}}^{i}(t+\tfrac{1}{2} {\vartriangle} t) \; ,\\
\vb*{\ddot{u}}^{i}(t+{\vartriangle} t) &= 1/\rho \, \left(\vb*{b}_0^{\mathrm{int}, \, i}(t+{\vartriangle} t) + \vb*{b}_0^{\mathrm{ext}, \, i}(t+{\vartriangle} t)\right) \; ,\\
\vb*{\dot{u}}^{i}(t+{\vartriangle} t) &= \vb*{\dot{u}}^{i}(t+\tfrac{1}{2} {\vartriangle} t) + \frac{{\vartriangle} t}{2} \, \vb*{\ddot{u}}^{i}(t+{\vartriangle} t) \; ,
\end{align}
\end{subequations}
with the time step ${\vartriangle} t$.

\begin{figure}[th]
\centering
\includegraphics[]{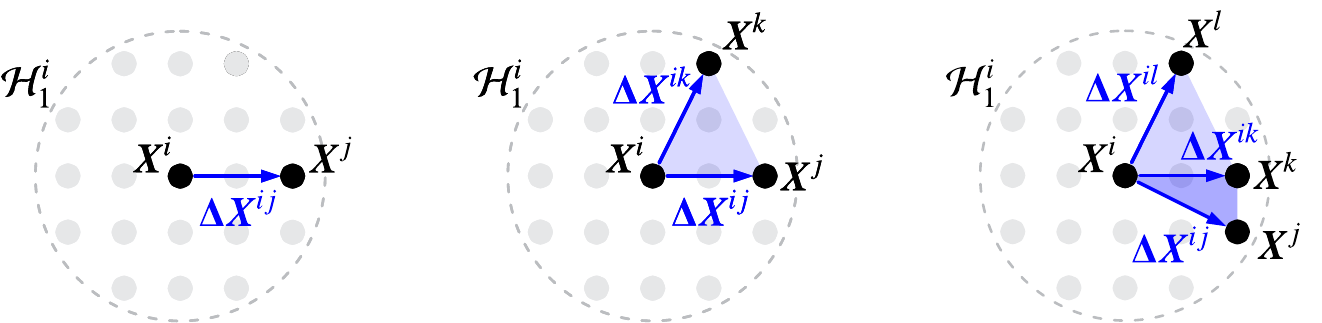}
\caption{Illustration of one-, two-, and three-neighbor interactions of point $\vb*{X}^i$}
\label{fig:one_two_three_ni}
\end{figure}

Various formulations of peridynamics exist for the calculation of the internal force density, and all of them are based on the non-local interactions between the material points.
The unique point of continuum-kinematics-based peridynamics is the use of three different types of interactions, also called one-, two- and three-neighbor interactions (see \autoref{fig:one_two_three_ni}).
Correspondingly, for continuum-kinematics-based peridynamics, $\vb*{b}_0^{\mathrm{int}, \, i}$ is the sum of the internal force densities of these interactions, thus
\begin{equation}
\vb*{b}_0^{\mathrm{int}, \, i} = \vb*{b}_1^{\mathrm{int}, \, i} + \vb*{b}_2^{\mathrm{int}, \, i} + \vb*{b}_3^{\mathrm{int}, \, i} \; .
\end{equation}

\subsection{One-neighbor interactions}

The \emph{one-neighbor interaction} of point $i$ and $j$, in standard peridynamics also called the \emph{bond}, is defined in material and current configuration as
\begin{align}
\vb*{\Delta X}^{ij} &= \vb*{X}^j - \vb*{X}^i \; , &
\vb*{\Delta x}^{ij} &= \vb*{x}^j - \vb*{x}^i \; .
\end{align}
One-neighbor interactions can be interpreted as line elements with the initial length $L^{ij}$ in material notation and the deformed length $l^{ij}$ in current configuration.
These so called relative length measures of the one-neighbor interaction are defined as
\begin{align}
L^{ij} &= \abs{\vb*{\Delta X}^{ij}} \; ,\qquad
l^{ij} = \abs{\vb*{\Delta x}^{ij}} \; .
\end{align}
It is assumed, that all one-neighbor interactions of point $i$ contribute equally. Therefore, an effective one-neighbor volume is defined as
\begin{equation}
V_1^{i} = \frac{V_{\mathcal{H}}^i}{N_1^i}
\end{equation}
with $N_1^i$ being the number of one-neighbor interactions for point $i$ and the neighborhood volume 
\begin{equation}
V_{\mathcal{H}}^i = \left\{ \begin{array}{ll}
    \beta^i \, \frac{4}{3} \, \pi \, \delta^3 & \quad \text{(3D problems)}\\
    \beta^i \, \pi \, \delta^2 & \quad \text{(2D problems)}
\end{array}
\right.
\end{equation}
with the factor $\beta^i \in [0,1]$ that takes the fullness of the neighborhood into account.
As an example, it applies $\beta^i = 1$ if the neighborhood of point $i$ is completely inside the body $\mathcal{B}_0$.
In contrast, if the neighborhood of point $i$ is partially outside the body $\mathcal{B}_0$, the factor $\beta^i < 1$ works as a correction factor to the volume $V_{\mathcal{H}}^i$.

The force density due to one-neighbor interactions is defined as 
\begin{equation}
\vb*{b}_{1}^{\mathrm{int}, \, i} = \int_{\mathcal{H}_1^i} \frac{\partial \psi_1^{ij}}{\partial \vb*{\Delta x}^{ij}} \; \mathrm{d} V_1^i
\end{equation}
with the harmonic potential
\begin{equation}
\psi_1^{ij} = \frac{1}{2} C_1 L^{ij} \left(\frac{l^{ij}}{L^{ij}} - 1\right)^2
\end{equation}
and the one-neighbor interaction constant $C_1$.
The constant $C_1$ can be interpreted as a resistance against the length change of one-neighbor interactions.
With 
\begin{equation}
\frac{\partial \psi_1^{ij}}{\partial \vb*{\Delta x}^{ij}} = C_1 \left( \frac{1}{L^{ij}} - \frac{1}{l^{ij}} \right) \vb*{\Delta x}^{ij} 
\end{equation}
the internal force density $\vb*{b}_{1}^{\mathrm{int}, \, i}$ of point $i$ can be formulated as
\begin{equation}\label{eq:b_int_1}
\vb*{b}_{1}^{\mathrm{int}, \, i} = \int_{\mathcal{H}_1^i} C_1 \left( \frac{1}{L^{ij}} - \frac{1}{l^{ij}} \right) \vb*{\Delta x}^{ij} \; \mathrm{d} V_1^i \; .
\end{equation}

\subsection{Two-neighbor interactions}

\begin{figure}[ht]
\centering
\includegraphics[]{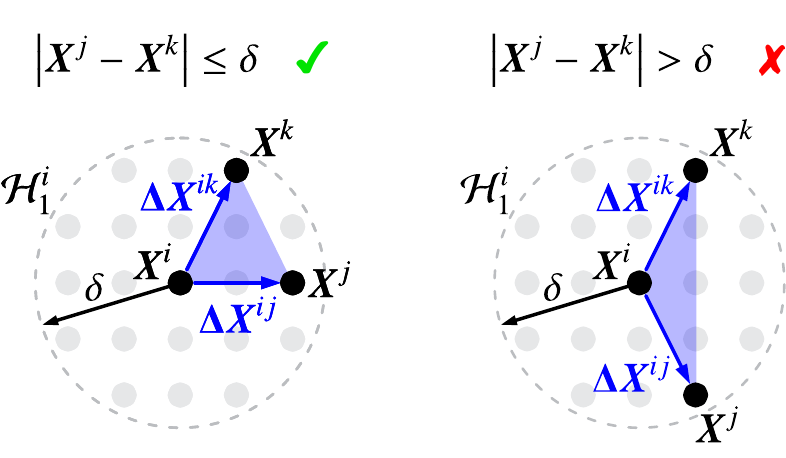}
\caption{Valid and invalid two-neighbor interaction of point $i$}
\label{fig:valid_invalid_twoni}
\end{figure}

Two-neighbor interactions are area elements, respectively triangles, spanned by the points $\vb*{X}^i$, $\vb*{X}^j$ and $\vb*{X}^k$.
They are constructed by two corresponding one-neighbor interactions $\vb*{\Delta X}^{ij}$ and $\vb*{\Delta X}^{ik}$ of point~$i$.
One important condition is that the distance between the points $\vb*{X}^j$ and $\vb*{X}^k$ needs to be smaller than the horizon $\delta$, as displayed in \autoref{fig:valid_invalid_twoni}.
Therefore, the set of all corresponding point-sets for two-neighbor interactions of point $i$ is defined as
\begin{equation}
  \mathcal{H}_2^i = \left\{ \big( \vb*{X}^j,\vb*{X}^k \big)
\in \mathcal{H}_1^i \times \mathcal{H}_1^i 
  \; | \;  0 < \abs{\vb*{X}^j - \vb*{X}^k} \leq \delta \right\} \quad \forall \; \vb*{X}^i \in \mathcal{B}_0 \; .
\end{equation}

The deformation of two-neighbor interactions is mainly described by the relative area measure, in material and current notation defined as
\begin{align}
  \vb*{A}^{ijk} &= \vb*{\Delta X}^{ij} \times \vb*{\Delta X}^{ik} \; ,
  \qquad
\vb*{a}^{ijk} = \vb*{\Delta x}^{ij} \times \vb*{\Delta x}^{ik} \; ,
\end{align}
and as scalar quantities the areas
\begin{align}
A^{ijk} &= \abs{\vb*{A}^{ijk}} \; ,\qquad
a^{ijk} = \abs{\vb*{a}^{ijk}} \; .
\end{align}
The force density due to two-neighbor interactions is defined as 
\begin{equation}
\vb*{b}_{2}^{\mathrm{int}, \, i} = \int_{\mathcal{H}_2^i} 2 \vb*{\Delta x}^{ik} \times \frac{\partial \psi_2^{ijk}}{\partial \vb*{a}^{ijk}} \; \mathrm{d} V_2^i
\end{equation}
with the harmonic potential
\begin{equation}
\psi_2^{ijk} = \frac{1}{2} C_2 A^{ijk} \left(\frac{a^{ijk}}{A^{ijk}} - 1\right)^2 \; ,
\end{equation}
and the effective two-neighbor volume
\begin{equation}
V_2^{i} = \frac{\left(V_{\mathcal{H}}^i\right)^2}{N_2^i} \; .
\end{equation}
The number of two-neighbor interactions of point $i$ is $N_2^i$.
The two-neighbor interaction constant $C_2$ can be interpreted as a resistance against the area change.
With
\begin{equation}
\frac{\partial \psi_2^{ijk}}{\partial \vb*{a}^{ijk}} = C_2 \left( \frac{1}{A^{ijk}} - \frac{1}{a^{ijk}} \right) \vb*{a}^{ijk} \; ,
\end{equation}
the internal force density $\vb*{b}_{2}^{\mathrm{int}, \, i}$ of point $i$ can be formulated as
\begin{equation}\label{eq:b_int_2}
\vb*{b}_{2}^{\mathrm{int}, \, i} = \int_{\mathcal{H}_2^i} 2 \, C_2 \, \vb*{\Delta x}^{ik} \times \left( \frac{1}{A^{ijk}} - \frac{1}{a^{ijk}} \right) \vb*{a}^{ijk} \; \mathrm{d} V_2^i \; .
\end{equation}

\subsection{Three-neighbor interactions}

Three-neighbor interactions are volume elements, precisely tetrahedrons, spanned by the points $\vb*{X}^i$, $\vb*{X}^j$, $\vb*{X}^k$ and $\vb*{X}^l$.
They are constructed by the three corresponding one-neighbor interactions $\vb*{\Delta X}^{ij}$, $\vb*{\Delta X}^{ik}$ and $\vb*{\Delta X}^{il}$ of point $i$.
For a valid three-neighbor interaction, the conditions
\begin{align}
  0 < \abs{\vb*{X}^j - \vb*{X}^k} \leq \delta \; ,\qquad
0 < \abs{\vb*{X}^j - \vb*{X}^l} \leq \delta \; ,\qquad
0 < \abs{\vb*{X}^k - \vb*{X}^l} \leq \delta \; ,
\end{align}
must be met.
Consequently, the set of all corresponding point-sets for three-neighbor interactions of point $i$ is defined as
\begin{align}
  \mathcal{H}_3^i = \left\{ \big( \vb*{X}^j,\vb*{X}^k,\vb*{X}^l \big)
\in \mathcal{H}_1^i\times \mathcal{H}_1^i\times \mathcal{H}_1^i
  \; | \;  \right. &0 < \abs{\vb*{X}^j - \vb*{X}^k} \leq \delta \; , \nonumber \\
&0 < \abs{\vb*{X}^j - \vb*{X}^l} \leq \delta \; , \nonumber\\
&0 \left.< \abs{\vb*{X}^k - \vb*{X}^l} \leq \delta \right\} \quad \forall \; \vb*{X}^i \in \mathcal{B}_0 \; .
\end{align}

The deformation of three-neighbor interactions is mainly described by the relative volume measure, in material and current notation defined as
\begin{align}
V^{ijkl} &= \vb*{\Delta A}^{ijk} \cdot \vb*{\Delta X}^{il} \; ,\quad
v^{ijkl} = \vb*{\Delta a}^{ijk} \cdot \vb*{\Delta x}^{il} \; .
\end{align}
The force density due to three-neighbor interactions is defined as
\begin{equation}
\vb*{b}_{3}^{\mathrm{int}, \, i} = \int_{\mathcal{H}_3^i} 3 \left( \vb*{\Delta x}^{ik} \times \vb*{\Delta x}^{il} \right) \frac{\partial \psi_3^{ijkl}}{\partial v^{ijkl}} \; \mathrm{d} V_3^i
\end{equation}
with the harmonic potential
\begin{equation}
\psi_3^{ijkl} = \frac{1}{2} C_3 V^{ijkl} \left(\frac{v^{ijkl}}{V^{ijkl}} - 1\right)^2 \; , 
\end{equation}
and the effective three-neighbor volume
\begin{equation}
V_3^{i} = \frac{\left(V_{\mathcal{H}}^i\right)^3}{N_3^i} \; .
\end{equation}
The number of three-neighbor interactions of point $i$ is $N_3^i$.
The three-neighbor interaction constant $C_3$ can be interpreted as a resistance against the volume change.
With
\begin{equation}
\frac{\partial \psi_3^{ijkl}}{\partial v^{ijkl}} = C_3 \left( \frac{1}{\abs{V^{ijkl}}} - \frac{1}{\abs{v^{ijkl}}} \right) v^{ijkl} \; ,
\end{equation}
the internal force density $\vb*{b}_{3}^{\mathrm{int}, \, i}$ of point $i$ can be formulated as
\begin{equation}\label{eq:b_int_3}
\vb*{b}_{3}^{\mathrm{int}, \, i} = \int_{\mathcal{H}_3^i} 3 \, C_3 \, \left( \vb*{\Delta x}^{ik} \times \vb*{\Delta x}^{il} \right) \left( \frac{1}{\abs{V^{ijkl}}} - \frac{1}{\abs{v^{ijkl}}} \right) v^{ijkl} \; \mathrm{d} V_3^i \; .
\end{equation}

\subsection{Contact}
For the modeling of contact between peridynamic bodies, we employ the algorithm by Silling and Askari \cite{silling2005}, which is in detail described in \cite{littlewood2015}.
The approach of this algorithm is mainly based on short-range forces that are activated at a certain threshold of the point distance.

For the incorporation of contact, Eqn.~\eqref{eq:equation_of_motion} needs to be extended to
\begin{equation}
\rho \, \vb*{\ddot{u}}^{i} = \vb*{b}_0^{\mathrm{int}, \, i} + \vb*{b}_0^{\mathrm{con}, \, i} + \vb*{b}_0^{\mathrm{ext}, \, i} \qquad \forall \; \vb*{X}^i \in \mathcal{B}_t, \; t \geq 0
\end{equation}
with the contact force density $\vb*{b}_0^{\mathrm{con}, \, i}$.
We consider two different peridynamic bodies $\mathcal{B}^I$ and $\mathcal{B}^{II}$, discretized with the point spacings $\vb*{\Delta x_I}$ and $\vb*{\Delta x_{II}}$.
Then we define the contact point sets $\mathcal{H}_{t}^{\mathrm{con}, \, i}$ and $\mathcal{H}_{t}^{\mathrm{con}, \, j}$ as
\begin{align}
\mathcal{H}_{t}^{\mathrm{con}, \, i} &= \left\{ \vb*{x}^j \in \mathcal{B}_t^{II} \; | \; 0 < \abs{\vb*{x}^{j} - \vb*{x}^{i}} \leq l_c \right\} \quad \forall \; \vb*{x}^i \in \mathcal{B}_t^{I}\\
\mathcal{H}_{t}^{\mathrm{con}, \, j} &= \left\{ \vb*{x}^i \in \mathcal{B}_t^{I} \; | \; 0 < \abs{\vb*{x}^{i} - \vb*{x}^{j}} \leq l_c \right\} \quad \forall \; \vb*{x}^j \in \mathcal{B}_t^{II}
\end{align}
with the critical contact distance $l_c \approx \mathrm{max}\left(\frac{\vb*{\Delta x_I}}{2}, \, \frac{\vb*{\Delta x_{II}}}{2}\right)$.
Now, the contact force densities yield to
\begin{align}
\vb*{b}_0^{\mathrm{con}, \, i} &= \int_{\mathcal{H}_{t}^{\mathrm{con}, \, i}} \frac{9 \, C^{\mathrm{con}}}{\pi \, \delta^5} \left(l_c - \abs{\vb*{x}^j - \vb*{x}^i}\right) \cdot \frac{\vb*{x}^j - \vb*{x}^i}{\abs{\vb*{x}^j - \vb*{x}^i}} \; \mathrm{d} V^j \quad \forall \; \vb*{x}^i \in \mathcal{B}_t^{I}\\
\vb*{b}_0^{\mathrm{con}, \, j} &= \int_{\mathcal{H}_{t}^{\mathrm{con}, \, j}} \frac{9 \, C^{\mathrm{con}}}{\pi \, \delta^5} \left(l_c - \abs{\vb*{x}^i - \vb*{x}^j}\right) \cdot \frac{\vb*{x}^i - \vb*{x}^j}{\abs{\vb*{x}^i - \vb*{x}^j}} \; \mathrm{d} V^i \quad \forall \; \vb*{x}^j \in \mathcal{B}_t^{II}
\end{align}
with the contact spring constant $C^{\mathrm{con}}$ and the point volumes $V^i$ and $V^j$.

\section{Modeling damage with continuum-kinematics-based peridynamics}

In classical peridynamics, damage is modeled by the failure of one-neighbor interactions.
The failure quantity for the strain-based damage model reads
\begin{equation}
d_1^{ij} = \left\{\begin{array}{ll}
0 & \text{if } \varepsilon^{ij} > \varepsilon_c\\
1 & \text{else}
\end{array}
\right.
\end{equation}
with the one-neighbor interaction stretch
\begin{equation}
\varepsilon^{ij} = \frac{\abs{\vb*{\Delta x}^{ij} - \vb*{\Delta X}^{ij} }}{\abs{\vb*{\Delta X}^{ij}}} \; ,
\end{equation}
and the critical one-neighbor interaction stretch $\varepsilon_c$.
This stretch can assumed to be identical to the critical bond stretch in classical peridynamics, which has been estimated by Madenci and Oterkus \cite{madenci2014} as
\begin{equation}
\displaystyle
\varepsilon_c = \left\{ \begin{array}{ll}
    \displaystyle \sqrt{\frac{G_c}{\delta \, \left( 3G + \left(\frac{3}{4}\right)^4 \left(K - \frac{5 G}{3}\right) \right)}}  & \quad \text{(3D problems)}\\[9mm]
    \displaystyle \sqrt{\frac{G_c}{\delta \, \left( \frac{6}{\pi} G + \frac{16}{9 \pi^2} \left(K - 2G\right) \right)}} & \quad \text{(2D problems)}\\
\end{array}
\right.
\end{equation}
with the critical energy release rate $G_c$, the shear modulus $G$ and the bulk modulus $K$.
The pointwise damage quantity $D^i$ incorporates the whole neighborhood, and is defined as
\begin{equation}
  D^{i} = 1 - \frac{\int_{\mathcal{H}_1^i} d_1^{ij} \; \mathrm{d} V_1^i}
  {\int_{\mathcal{H}_1^i} \mathrm{d} V_1^i}
  \,.
\end{equation}
%
These equations cannot directly be used to model damage within the continuum-kinematics-based framework, because they do not take two- or three-neighbor interactions into consideration.
Applying this damage model alone will not lead to crack paths but to diffuse failure zones, because two- or three-neighbor interactions are still active and lead to forces between failed points.

To adress this problem, failure quantities for two- and three-neighbor interactions, $d_2^{ijk}$ and $d_3^{ijkl}$, are introduced.
Here we propose that two- and three-neighbor interactions fail, if one or more corresponding one-neighbor interactions fail.
Therefore, the failure quantity for two-neighbor interactions can be defined as
\begin{equation}
d_2^{ijk} = \left\{\begin{array}{ll}
    0 & \text{if } d_1^{ij}= 0 \text{ or } d_1^{ik} = 0 \; ,\\
    1 & \text{else} \; ,
\end{array}
\right .
\end{equation}
and for three-neighbor interactions as
\begin{equation}
d_3^{ijkl} = \left\{\begin{array}{ll}
    0 & \text{if } d_1^{ij} = 0\text{ or } d_1^{ik}= 0\text{ or } d_1^{il} = 0 \; ,\\
    1 & \text{else} \; .
\end{array}
\right .
\end{equation}

With these failure quantities, we re-define the internal force density for one-neighbor interactions \eqref{eq:b_int_1} as
\begin{equation}
\vb*{b}_{1}^{\mathrm{int}, \, i} = \int_{\mathcal{H}_1^i} d_1^{ij} \, C_1 \left( \frac{1}{L^{ij}} - \frac{1}{l^{ij}} \right) \vb*{\Delta x}^{ij} \; \mathrm{d} V_1^i \; ,
\end{equation}
for two-neighbor interactions \eqref{eq:b_int_2} as
\begin{equation}
\vb*{b}_{2}^{\mathrm{int}, \, i} = \int_{\mathcal{H}_2^i} d_2^{ijk} \, 2 \, C_2 \, \vb*{\Delta x}^{ik} \times \left( \frac{1}{A^{ijk}} - \frac{1}{a^{ijk}} \right) \vb*{a}^{ijk} \; \mathrm{d} V_2^i \; ,
\end{equation}
and for three-neighbor interactions \eqref{eq:b_int_3} as
\begin{equation}
\vb*{b}_{3}^{\mathrm{int}, \, i} = \int_{\mathcal{H}_3^i} d_3^{ijkl} \, 3 \, C_3 \, \left( \vb*{\Delta x}^{ik} \times \vb*{\Delta x}^{il} \right) \left( \frac{1}{\abs{V^{ijkl}}} - \frac{1}{\abs{v^{ijkl}}} \right) v^{ijkl} \; \mathrm{d} V_3^i \; .
\end{equation}
In such a manner, the failed point interactions do not contribute to the internal material response and their damaging effect is considered.

\section{Numerical results for the mode I tension test}

\begin{figure}[ht]
\centering
\includegraphics[]{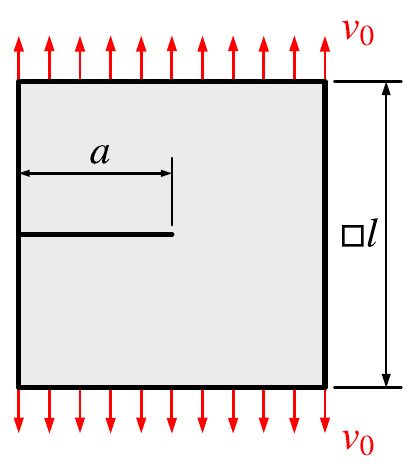}%
\caption{Setup of the mode I tension test}
\label{fig:tensiontest_setup}
\end{figure}

In the following section, it is shown, that continuum-kinematics-based peridynamics is able to model crack growth for two- and three-dimensional problems.
Therefore, a square with edge length $l$ and a predefined crack of length $a=\frac{1}{2}l$ is subjected to tension due to the expansion of the upper and lower region of the model with a constant velocity $v_0=\SI{0.005}{\metre\per\second}$ (see \autoref{fig:tensiontest_setup}).

\begin{table}[ht]
\begin{center}
\caption{Parameters for calculations of the mode I tension test}
\label{tab:param_mode_I}
\begin{tabular}{lcc} \hline
Parameter & 2D setup & 3D setup\\\hline
Density $\rho$ & \SI{7580}{\kilogram\per\metre\squared} & \SI{7580}{\kilogram\per\cubic\metre} \\
Poisson's ratio $\nu$ & \num{0.3} & \num{0.3} \\
Young's modulus $E$ & \SI{210000}{MPa} & \SI{210000}{MPa} \\
Griffith's parameter $G_c$ & \SI{140}{\newton\per\metre} & \SI{500}{\newton\per\metre} \\
Point spacing $\Delta x$ & \SI{5}{mm} & \SI{16.7}{mm} \\
Horizon $\delta$ & \SI{15.075}{mm} & \SI{50.25}{mm} \\\hline
\end{tabular}
\end{center}
\end{table}

For the 2D setup, a uniformly distributed point cloud with $200 \times 200$ points, and for the 3D setup $60 \times 60 \times 3$ points are used.
Both setups use the material parameters of steel, as listed in \autoref{tab:param_mode_I}.
As derived by Ekiz, Javili, and Steinmann \cite{Ekiz2021}, the interaction constants of the two-dimensional setup are
\begin{align}\label{eq:cpd_param_2d}
C_1 &= \frac{12}{\pi \, \delta^3} \frac{E}{\nu + 1} \; , &
C_2 &= \frac{27}{16 \, \pi \, \delta^6} \frac{E \, (1 - 3 \, \nu)}{\nu^2 - 1} \; ,
\end{align}
with the Young's modulus $E$ and the Poisson's ratio $\nu$.
Furthermore, the constants for one- and three-neighbor interactions of the three-dimensional setup are defined as
\begin{align}\label{eq:cpd_param_3d}
C_1 &= \frac{30 \, \mu}{\pi \, \delta^4} \; , & 
C_3 &= \frac{32}{\pi^4 \, \delta^{12}} (\lambda-\mu) \; ,
\end{align}
for $C_2=0$, and with the first and second Lamé parameter $\lambda = \frac{E \, \nu}{(1 + \nu)(1-2\nu)}$ and $\mu = \frac{E}{2(1+\nu)}$ \cite{Ekiz2022}.

\begin{figure}[ht]
\centering
\includegraphics[height=7cm]{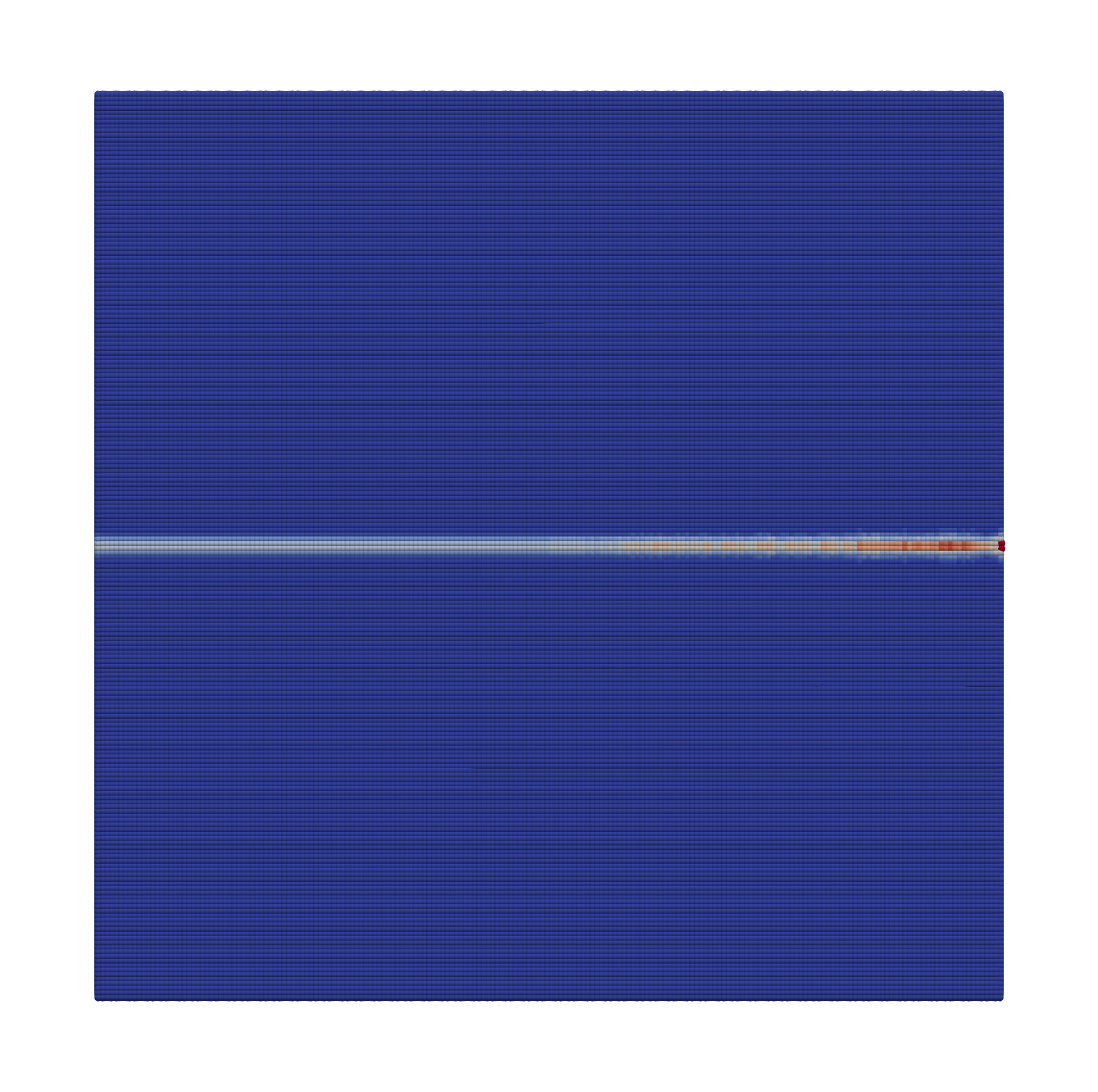}%
\includegraphics[height=7cm]{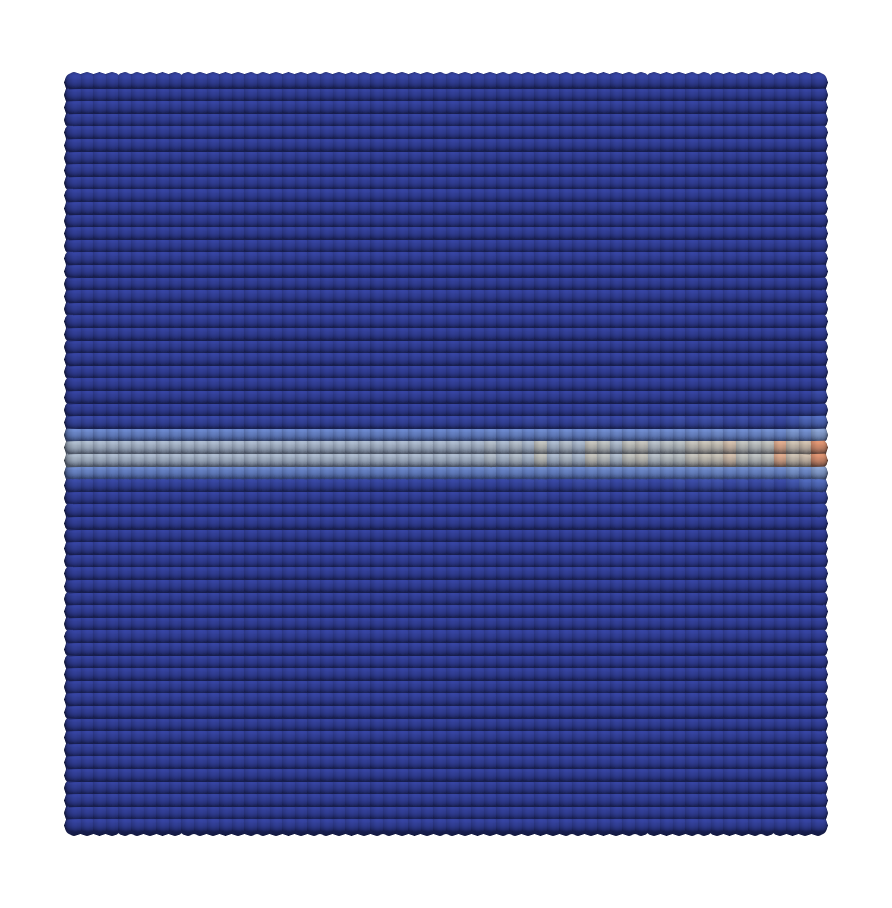}%
\includegraphics[]{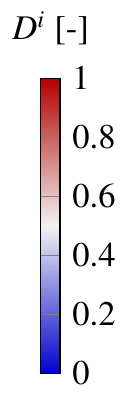}
\caption{Damage $D^i$ for the 2D (left) and 3D setup (right)}
\label{fig:tensiontest2D}
\end{figure}

In \autoref{fig:tensiontest2D}, the damage $D^i$ for the 2D and the 3D setup is shown.
The crack propagates and grows as expected until the square is broken into two pieces for both setups.
Without the additional failure quantities $d_2^{ijk}$ and $d_3^{ijkl}$, a diffuse damage field and no clear crack path would be the result of this simulations.
Consequently, continuum-kinematics-based peridynamics can be used to model crack-growth with our proposed damage model.

\section{Curved bar under pressure}

\begin{figure}[ht]
\centering
\includegraphics[]{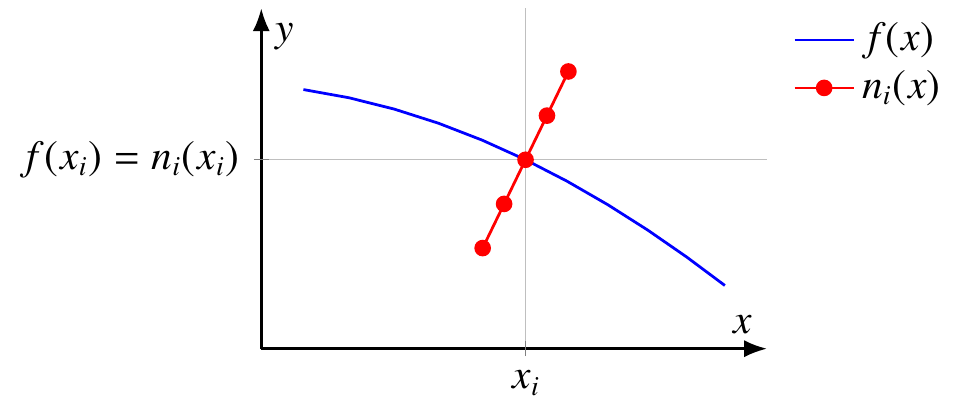}
\caption{Function $f(x)$ describing the curve of the bar and $n_i(x)$ with $N_n=5$ material points}
\label{fig:curves}
\end{figure}

\begin{figure}[ht]
\centering
\includegraphics[width=0.5\textwidth]{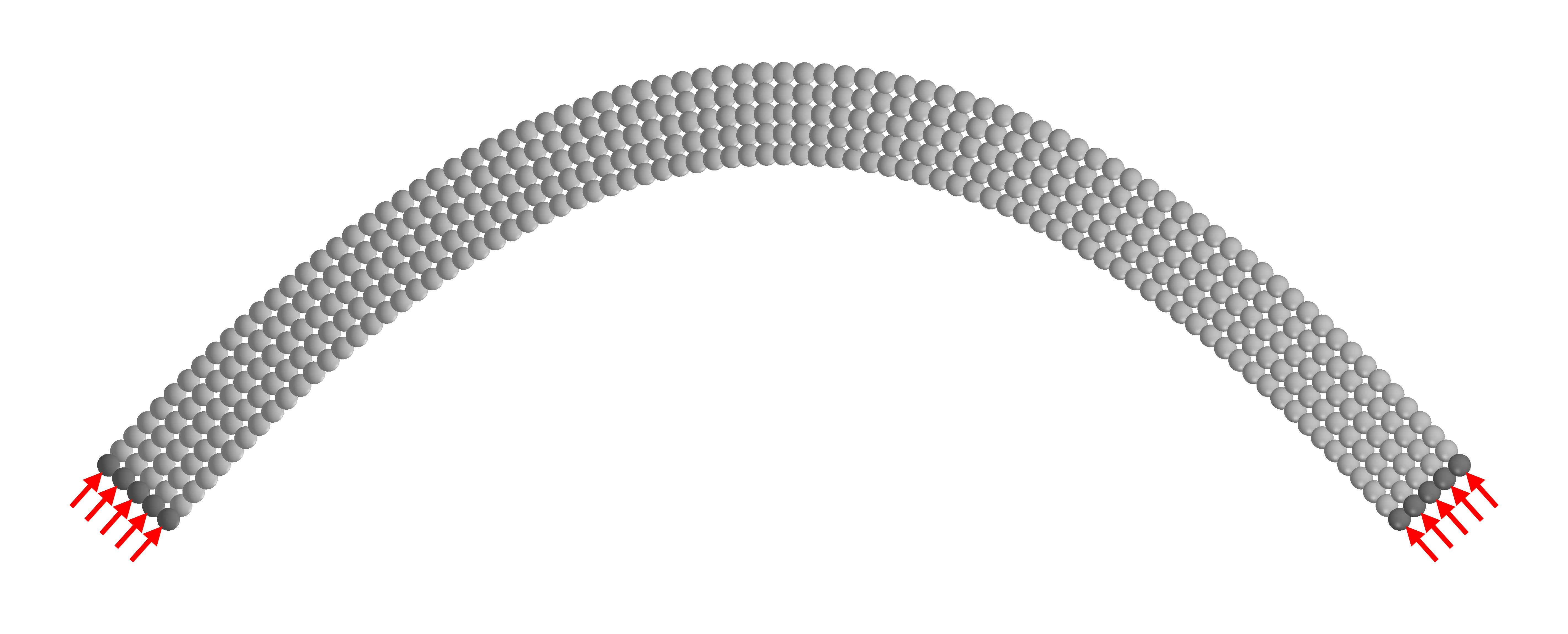}
\caption{Coarse point cloud with $N_n=5$ and one layer of points on each side used for a pressure impulse}
\label{fig:coarsecurvedbar}
\end{figure}

In the following section, crack initiation due to impact is investigated for a two- and three-di\-men\-sion\-al discretizations.
For this purpose, a model of a curved bar is subjected to pressure waves, which are supposed to superimpose inside the material and eventually lead to crack initiation.
The material points are spatially distributed along the curve $f(x)=\cos(\frac{\pi}{2} \, x)$ (see \cite{WeinbergWieners2022} for more details). 
As shown in \autoref{fig:curves}, for each root point $x_i$ on $f$, $N_n$ points occur along the curve
\begin{equation}
n_i(x) =  \frac{1}{f'(x_i)} \cdot (x-x_i) + f(x_i) \; ,
\end{equation}
with the derivative $f'(x) = \frac{\mathrm{d}f}{\mathrm{dx}} = - \frac{\pi}{2} \, \sin(\frac{\pi}{2} \, x)$.
The bar has the width $W_B = \SI{62.5}{mm}$ along the curve $n_i(x)$ and is defined for the root points $x_i \in [-\frac{L_B}{2}, \frac{L_B}{2}]$ with the bar length $L_B = \SI{1}{m}$.
The number of points $N_n$ on $n_i(x)$ is a measure to describe the density of the point cloud, since it is used to specify the point spacing $\Delta x = \frac{W_B}{N_n}$.
For the three-dimensional implementation, $N_n$ layers of material points are equally distributed with distance $\Delta x$ along $z \in [-\frac{W_B}{2}, \frac{W_B}{2}]$.
A coarse point cloud with $N_n=5$ is shown in \autoref{fig:coarsecurvedbar} for the purpose of illustrating the discretization.

\begin{figure}[ht]
\centering
\includegraphics[]{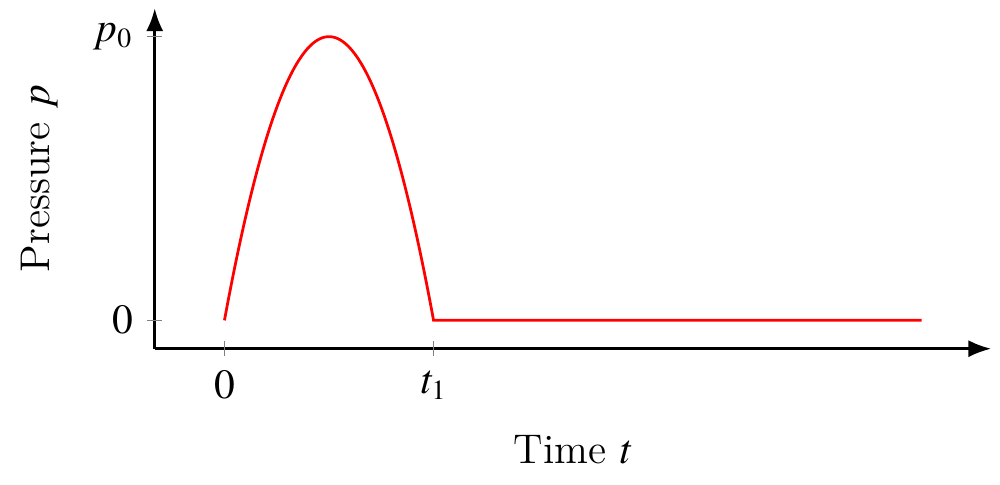}
\caption{Pressure impulse $p(t)$}
\label{fig:pressure_impulse}
\end{figure}

On each side of the curved bar, a pressure impulse
\begin{equation}
p(t) = - 4 \cdot \frac{p_0}{{t_1}^2} \cdot \left(t - \frac{t_1}{2}\right)^2 + p_0
\end{equation}
with the pressure peak $p_0$ and the impulse duration $t_1$ is applied for one layer of material points in the left and right boundary (see \autoref{fig:pressure_impulse}).
The pressure is applied via the external body force density
\begin{equation}
\vb*{b}^{\mathrm{ext}, \ i}_0 = \frac{p(t)}{\Delta x} \vb*{n}_{l/r}
\end{equation}
with the normal vector
\begin{equation}
\vb*{n}_l = \left\{ \begin{array}{ll}
\left[\sin{\alpha}, \; \cos{\alpha}\right]^{\mathrm{T}} & \quad \text{(2D problems)}\\
\left[\sin{\alpha}, \; \cos{\alpha}, \; 0\right]^{\mathrm{T}} & \quad \text{(3D problems)}\\
\end{array}
\right.
\end{equation}
for the left side and 
\begin{equation}
\vb*{n}_r = \left\{ \begin{array}{ll}
\left[-\sin{(\alpha)}, \; \cos{(\alpha)}\right]^{\mathrm{T}} & \quad \text{(2D problems)}\\
\left[-\sin{(\alpha)}, \; \cos{(\alpha)}, \; 0\right]^{\mathrm{T}} & \quad \text{(3D problems)}\\
\end{array}
\right.
\end{equation}
for the right side of the bar, and the angle $\alpha = \arctan{\left(-\frac{1}{f'(L_B/2)}\right)}$.

The material parameters used for the calculations are shown in \autoref{tab:matparam_curvedbar}.
For the two-di\-men\-sion\-al setup, a pressure impulse with the peak $p_0 = \SI{4e5}{\newton\per\metre}$ and for the three-dimensional setup, $p_0 = \SI{1e6}{\newton\per\metre\squared}$ is used.
For both setup's, the pulse has the duration $t_1 = \SI{300}{\micro\second}$.
Remark that for 2D, the body force density $\vb*{b}^{\mathrm{ext}}_0$ has the unit [$\si{\newton\per\metre\squared}$].
The interaction constants are calculated as before (see Eq.~(\ref{eq:cpd_param_2d}) and (\ref{eq:cpd_param_3d})).

\begin{table}[t]
\begin{center}
\caption{Parameters for calculations of the curved bar}
\label{tab:matparam_curvedbar}
\begin{tabular}{lcc} \hline
Parameter & 2D setup & 3D setup\\\hline
Density $\rho$ & \SI{7580}{\kilogram\per\metre\squared} & \SI{7580}{\kilogram\per\cubic\metre} \\
Poisson's ratio $\nu$ & \num{0.3} & \num{0.3} \\
Young's modulus $E$ & \SI{210000}{MPa} & \SI{210000}{MPa} \\
Griffith's parameter $G_c$ & \SI{1}{\newton\per\metre} & \SI{1}{\newton\per\metre} \\
Point spacing $\Delta x$ & \SI{3.125}{mm} & \SI{12.5}{mm} \\
Horizon $\delta$ & \SI{9.42}{mm} & \SI{38}{mm} \\\hline
\end{tabular}
\end{center}
\end{table}

\begin{figure}[!ht]
\centering
\begin{minipage}[b]{0.6\textwidth}
\includegraphics[width=\textwidth]{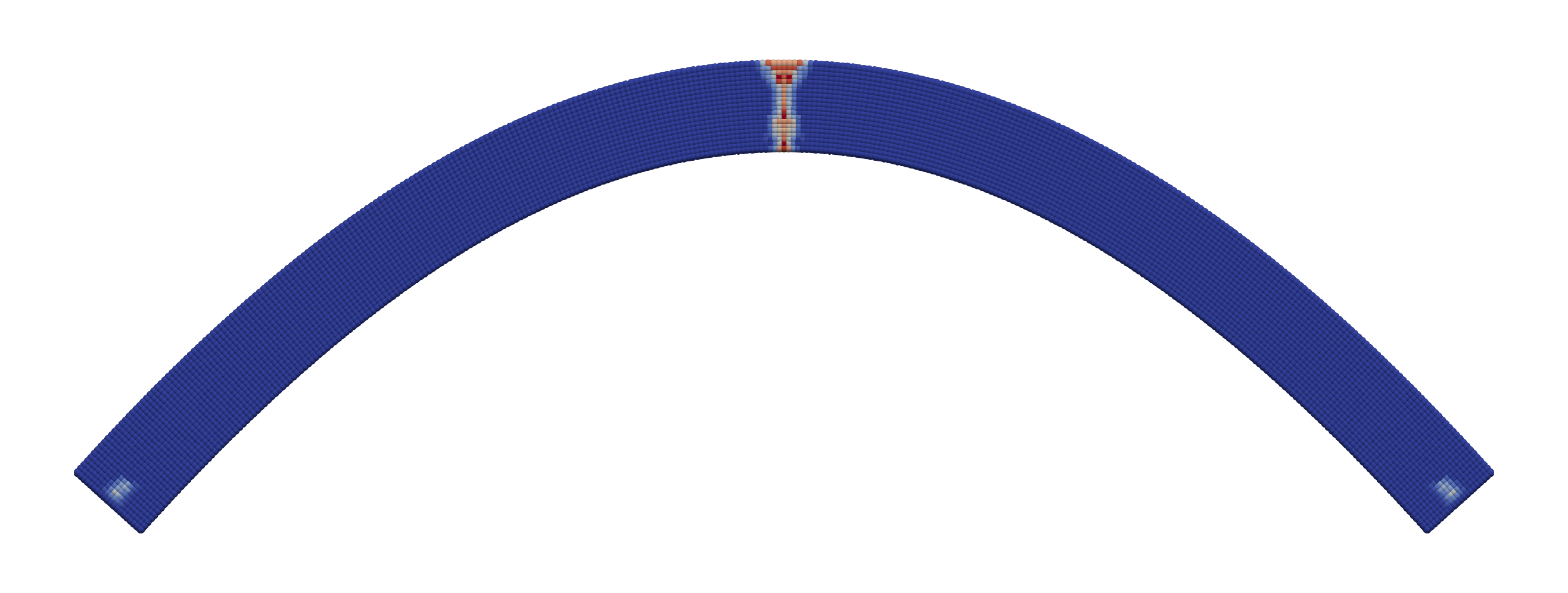}\\
\includegraphics[width=\textwidth]{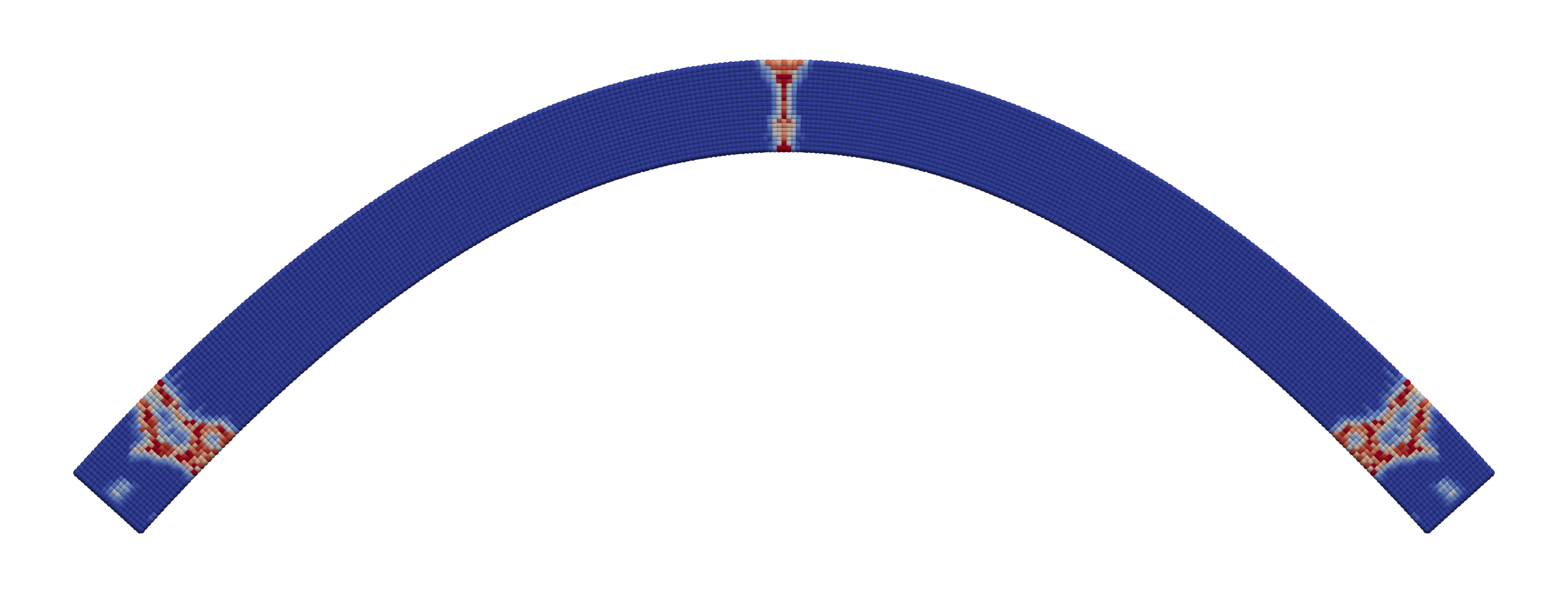}%
\end{minipage}%
\begin{minipage}[b]{0.1\textwidth}
\includegraphics[]{ParaViewColorbar.pdf}
\end{minipage}
\caption{Damage $D^i$ of the 2D curved bar for $t=\SI{2.9}{ms}$ (top) and $t=\SI{5.8}{ms}$ (bottom)}
\label{fig:res_curvedbar_2D}
\end{figure}

In \autoref{fig:res_curvedbar_2D}, the damage $D^i$ of the two-dimensional setup is shown.
After $t=\SI{2.9}{ms}$, a crack in the middle of the bar is visible.
The pressure waves propagate through the bar and then get reflected, which consequently transforms them into tensile waves.
These tensile waves then lead to the initiation of a crack.
The waves continue to propagate in the model and when superimposed again, the same effect occurs and more cracks are formed.
The two-dimensional model reproduces this behavior very well, since exactly these further cracks occur for time $t=\SI{5.8}{ms}$.

\begin{figure}[!t]
\centering
\includegraphics[width=0.6\textwidth]{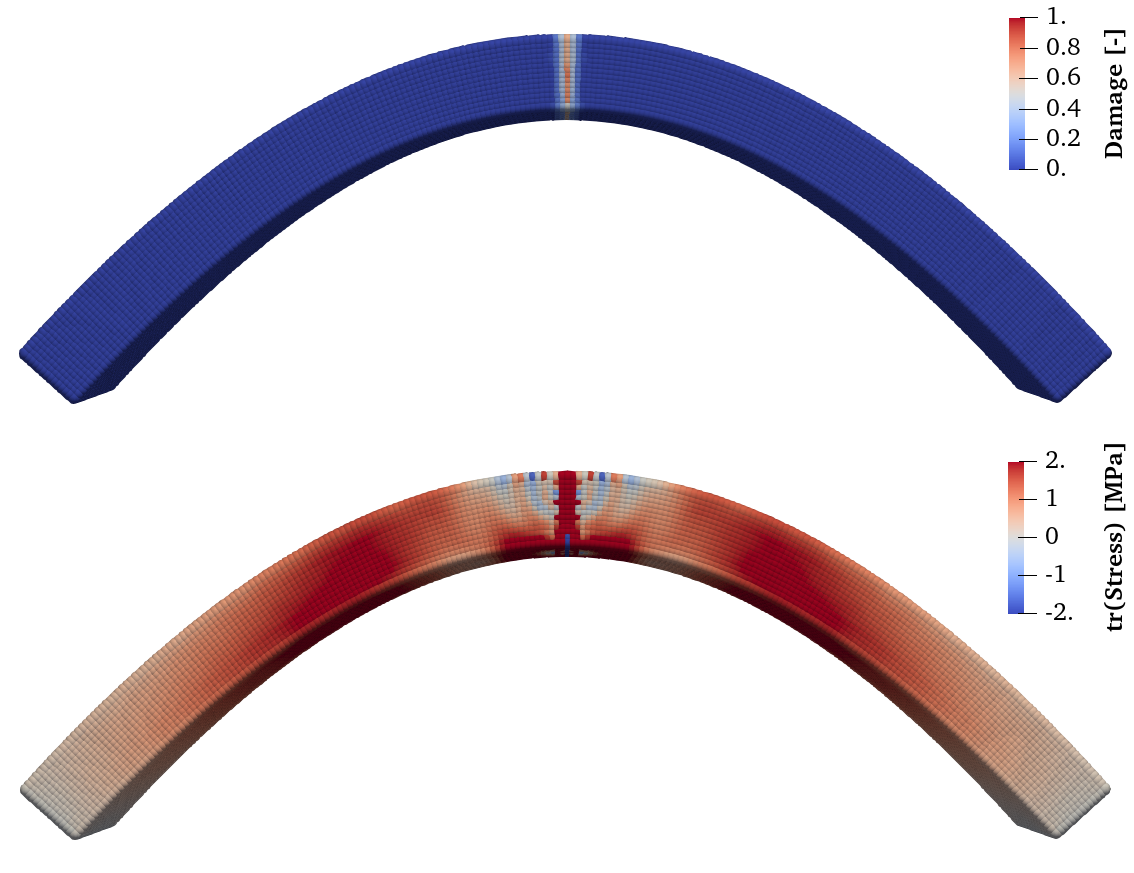}
\caption{Damage $D^i$ (top) and the trace of an artificially calculated stress tensor (bottom) of the 3D curved bar for $t=\SI{0.5}{ms}$ }
\label{fig:res_curvedbar_3D-1}
\end{figure}

\begin{figure}[!t]
\centering
\includegraphics[width=0.6\textwidth]{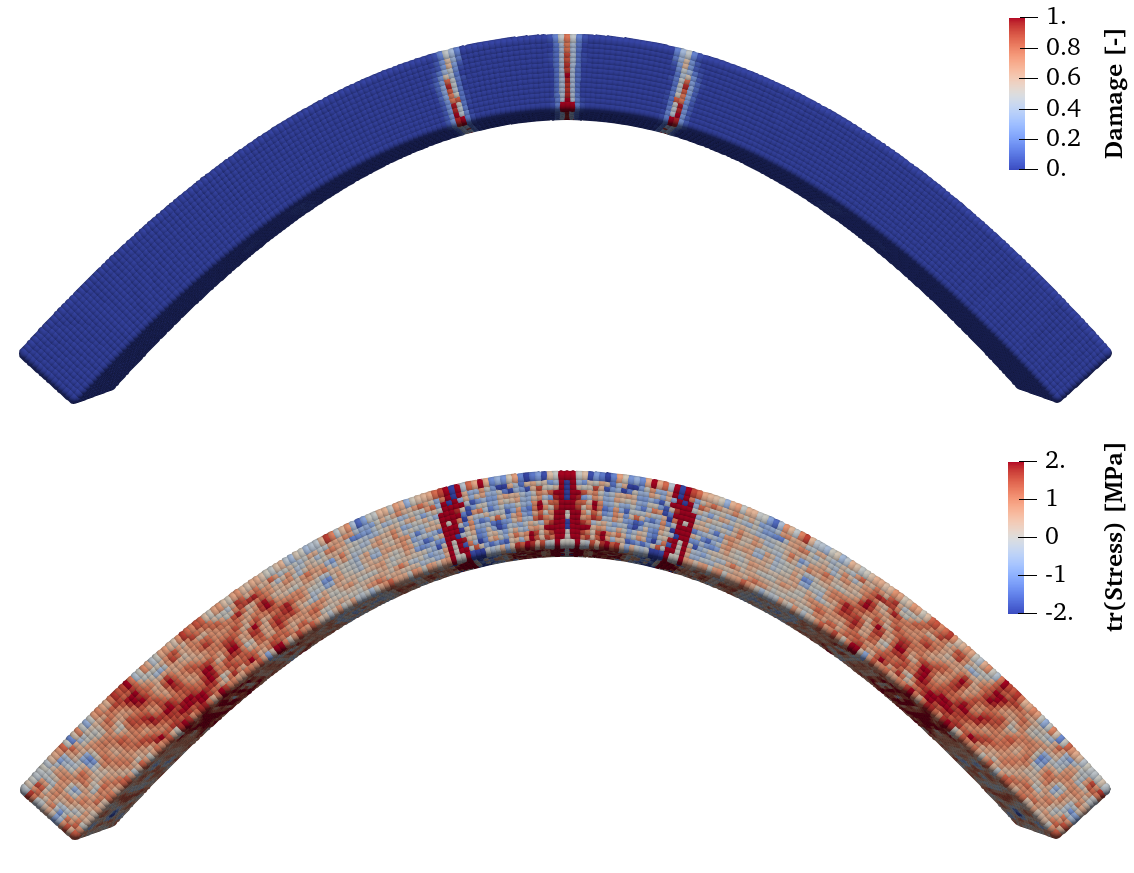}
\caption{Damage $D^i$ (top) and the trace of an artificially calculated stress tensor (bottom) of the 3D curved bar after $t_1=\SI{1.4}{ms}$ }
\label{fig:res_curvedbar_3D-2}
\end{figure}

The same behavior can also be observed with the 3D model (see \autoref{fig:res_curvedbar_3D-1}).
As an important remark, for the visualization of the waves in the model, a stress tensor was artificially calculated.
The calculation of the stresses is not part of the peridynamics and is only used to illustrate the wave reflection. 
After the first reflection of the pressure wave, a single crack is initiated in the center of the model.
Also the cracking due to the further superposition of the waves can be detected, as seen in \autoref{fig:res_curvedbar_3D-2}.
The position differs from that of the 2D model, but this could be explained by the versatile influencing factors of continuum-kinematics-based peridynamics, such as material parameters and different discretizations.
Here further studies are necessary.
In summary it can be stated, that continuum-kinematics-based peridynamics can be used to map cracking due to the material response to pressure waves.

\section{Impact damage}

\begin{figure}[!ht]
\centering
\includegraphics[]{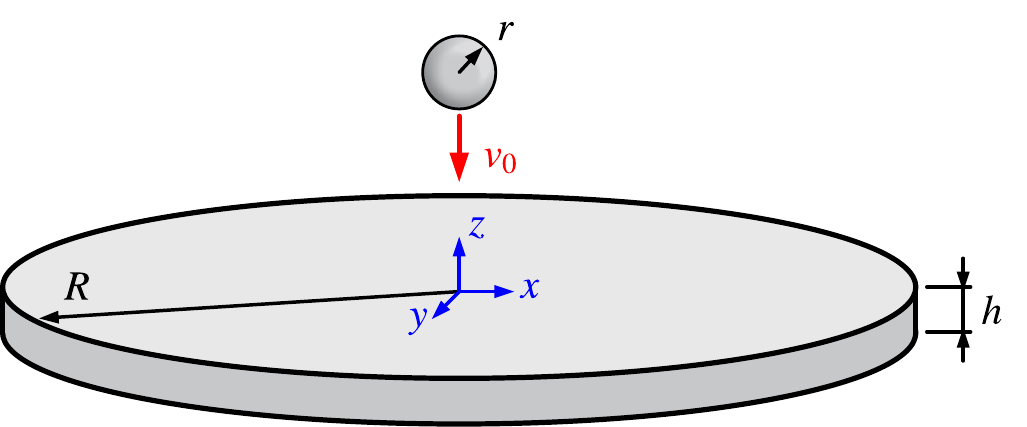}
\caption{Setup of the sphere and the circular disc}
\label{fig:impactmodel}
\end{figure}

In the following section, damage due to contact between two peridynamic bodies is investigated for our proposed damage model.
Here, the shot of a sphere with initial velocity $v_0 = \SI{50}{\metre\per\second}$ through a circular disc that is free in space is computed numerically.
As it is displayed in \autoref{fig:impactmodel}, the sphere has the radius $r=\SI{10}{mm}$ and the disc the radius $R=\SI{250}{mm}$ and height $h=\SI{10}{mm}$.
The material parameters for both bodies are listed in \autoref{tab:matparam_impact}.
For the modeling of the contact, the search radius $l_c = \SI{2.5}{mm}$ and the contact spring constant $C^{\mathrm{con}}=\SI{1000}{GPa}$ are used.

\begin{table}[ht]
\begin{center}
\caption{Parameters for the calculation of the sphere impact}
\label{tab:matparam_impact}
\begin{tabular}{lcc} \hline
Parameter & Sphere & Disc\\\hline
Density $\rho$ & \SI{7850}{\kilogram\per\metre\squared} & \SI{2000}{\kilogram\per\cubic\metre} \\
Poisson's ratio $\nu$ & \num{0.25} & \num{0.2} \\
Young's modulus $E$ & \SI{210000}{MPa} & \SI{50000}{MPa} \\
Griffith's parameter $G_c$ & \SI{1500}{\newton\per\metre} & \SI{1}{\newton\per\metre} \\
Point spacing $\Delta x$ & \SI{4}{mm} & \SI{5}{mm} \\
Horizon $\delta$ & \SI{12.06}{mm} & \SI{40.075}{mm} \\\hline
\end{tabular}
\end{center}
\end{table}

\begin{figure}[!ht]
\centering
\includegraphics[width=0.45\textwidth]{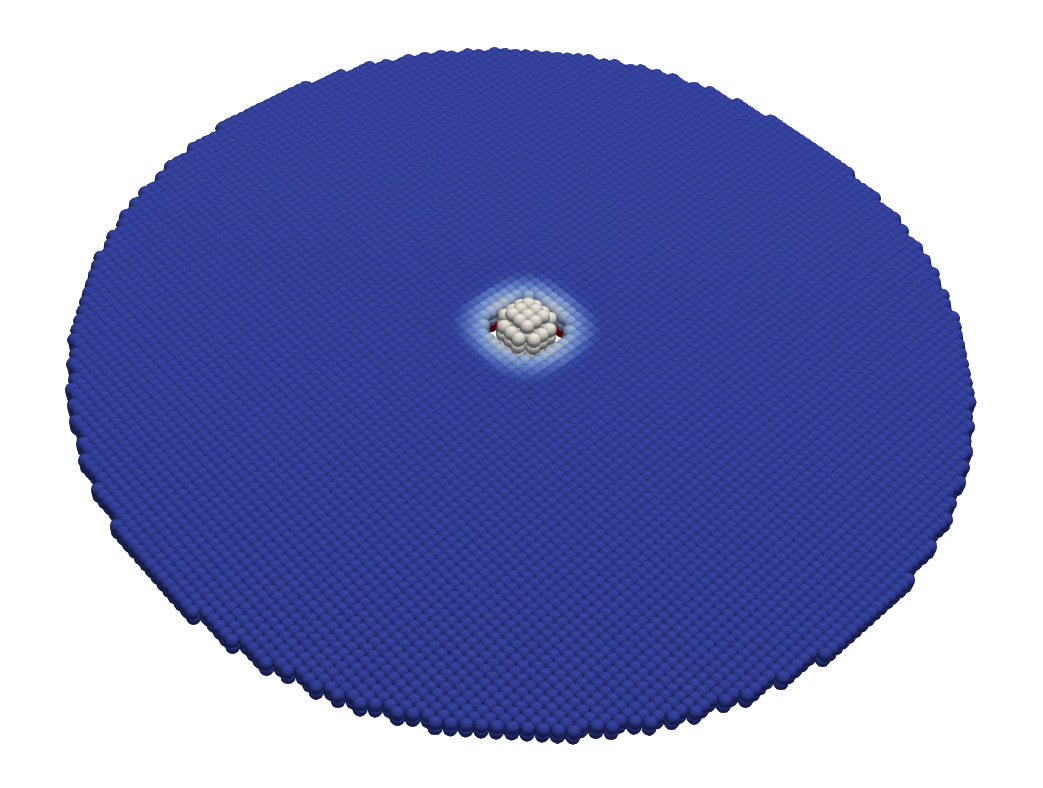}%
\includegraphics[width=0.45\textwidth]{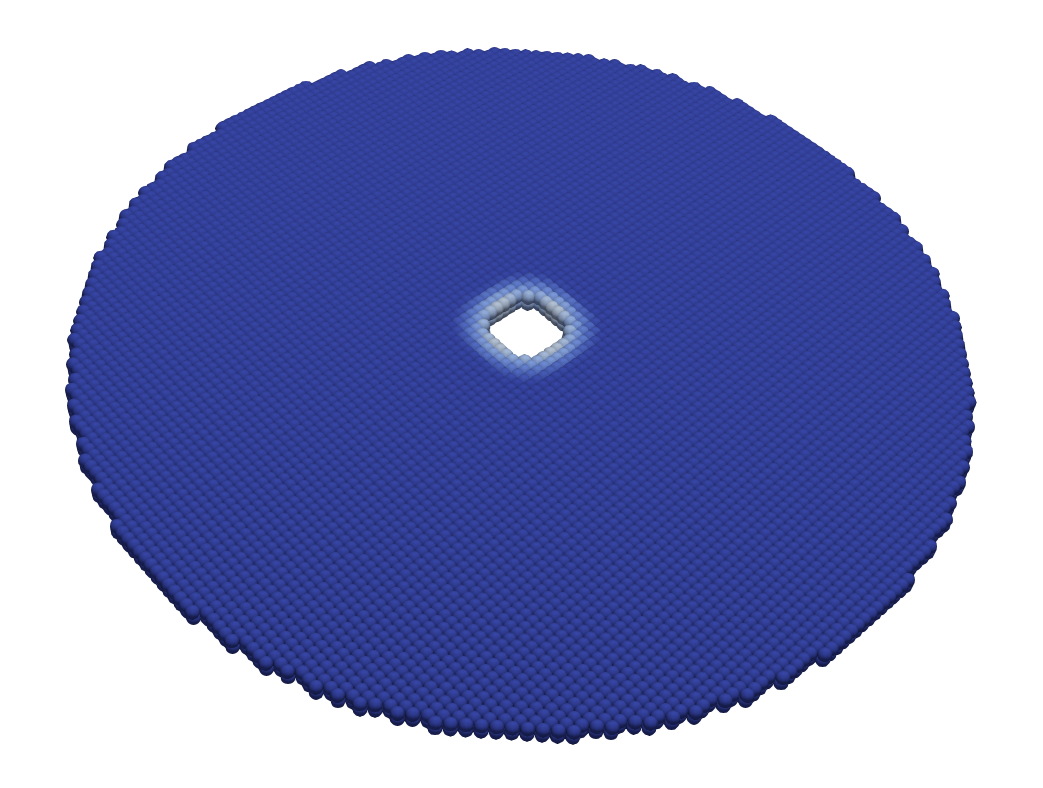}%
\includegraphics[]{ParaViewColorbar.pdf}
\caption{Damage $D^i$ for the time $t=\SI{0.26}{ms}$ (left) and $t=\SI{0.86}{ms}$ (right)}
\label{fig:impactresults}
\end{figure}

In \autoref{fig:impactresults}, the results of the numerical experiment are displayed for $t=\SI{0.26}{ms}$ (left) and $t=\SI{0.86}{ms}$ (right).
The sphere punches through the circular disc and stamps a hole in it.
Slight damage occurs in the impact area, but no further cracks propagate as a result of the impact.
Similar material response can be seen for variations of material or contact parameters.

\section{Summary}

In this study, we present an approach to dynamic fracture and impact damage with continuum-kinematic-based peridynamics.
We extend the classical damage model and introduce failure quantities for two- and three-neighbor interactions.
For two- and three-dimensional simulations, our approach handles crack growth of a mode I tension test very well.
We show that crack initiation due to the superposition of pressure waves agrees well with the expected results.
Additional impact simulations show, that continuum-kinematics-based peridynamics is even able to model a punch through without further cracking.

\section*{Acknowledgments}

The authors gratefully acknowledge the support of the Deutsche
Forschungsgemeinschaft (DFG)
in the projects \mbox{WE~2525/15-1} and \mbox{WI~1430/9-1}.

\section*{Conflict of interest}
The authors declare that they have no conflict of interest.


\end{document}